\documentclass[runningheads]{llncs}

\usepackage[T1]{fontenc}
\usepackage{graphicx}
\usepackage{subcaption} 
\usepackage{longtable}
\usepackage{wrapfig}
\usepackage{booktabs}
\usepackage{graphicx}  
\usepackage{booktabs}  
\usepackage{ragged2e}

\newcommand\blfootnote[1]{%
  \begingroup
  \renewcommand\thefootnote{}\footnote{#1}%
  \addtocounter{footnote}{-1}%
  \endgroup
}

\begin{document}
\title{Patient-Centred Explainability in IVF Outcome Prediction}
%
%
\author{Adarsa Sivaprasad\inst{1} \and
Ehud Reiter\inst{1} \and
David McLernon\inst{1}  \and Nava Tintarev\inst{2} \and Siladitya Bhattacharya\inst{1} \and Nir Oren\inst{1} }
\authorrunning{A. Sivaprasad et al.}
%
\institute{University of Aberdeen, Aberdeen, UK \email{\{a.sivaprasad.22@abdn.ac.uk}
\and Maastricht University, Maastricht, Netherlands}

\maketitle              
\begin{abstract}
This paper evaluates the user interface of an in vitro fertility (IVF) outcome prediction tool, focussing on its understandability for patients or potential patients. We analyse four years of anonymous patient feedback, followed by a user survey and interviews to quantify trust and understandability. Results highlight a lay user's need for prediction model \emph{explainability} beyond the model feature space. We identify user concerns about data shifts and model exclusions that impact trust. The results call attention to the shortcomings of current practices in explainable AI research and design and the need for explainability beyond model feature space and epistemic assumptions, particularly in high-stakes healthcare contexts where users gather extensive information and develop complex mental models. To address these challenges, we propose a dialogue-based interface and explore user expectations for personalised explanations.

\blfootnote{This preprint has not undergone peer review. The Version of Record of this contribution is accepted for presentation at AIIH 2025. }

\end{abstract}

\section{Introduction}

\begin{figure}[hbt!]
\centering
\includegraphics[width=0.95\linewidth]{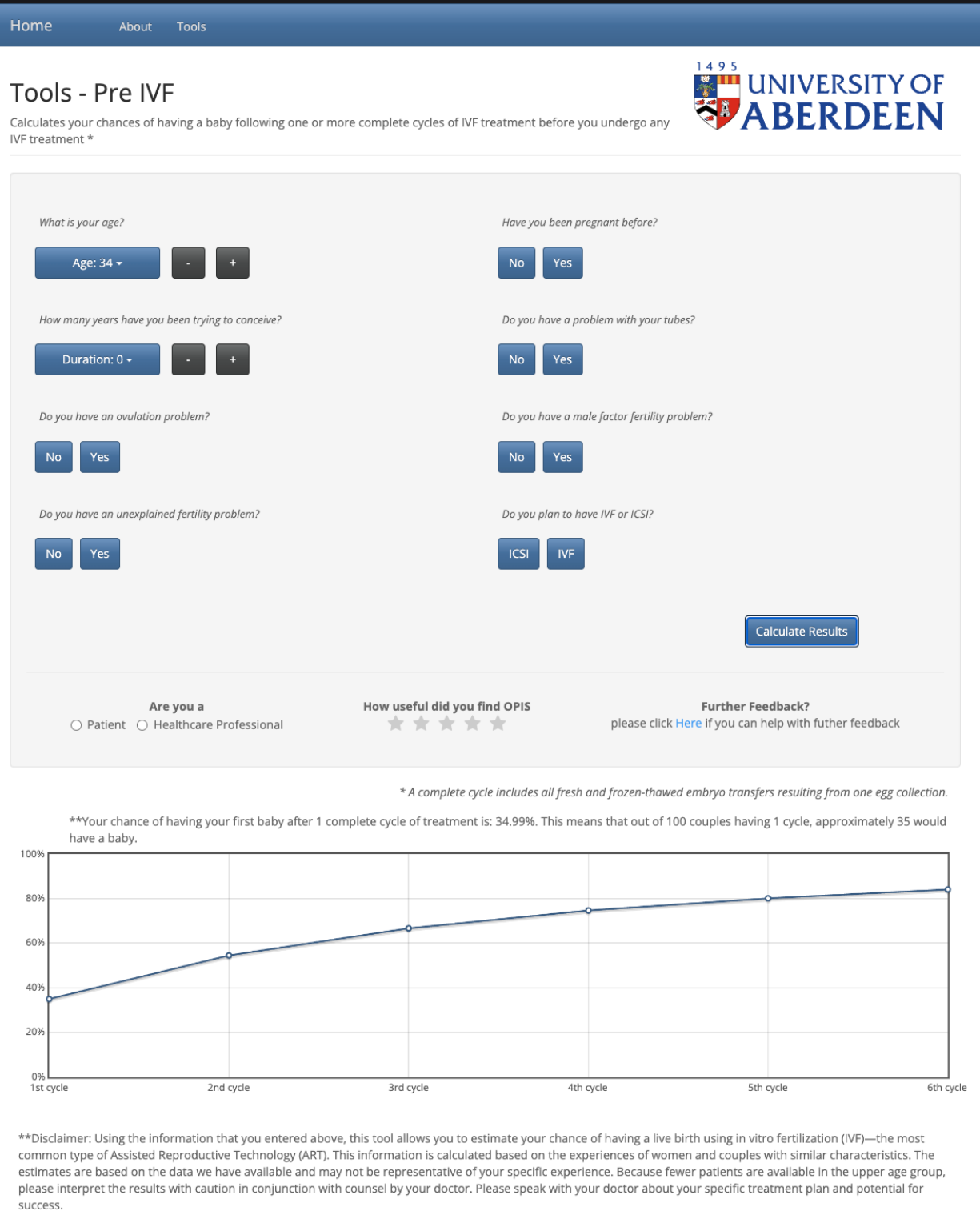}
\caption{A screenshot of the user interface of OPIS pre-IVF tool. }
\label{fig:opisUI}
\end{figure}

Research in explainable AI (XAI) focuses on model interpretability, uncertainty communication, and tailoring explanations to user needs \cite{doshi2017towards}, which are crucial for trust, accountability, and their usability\cite{samek2017explainable}. Although there have been discussions on social science aspects of understanding probabilistic reasoning for lay users\cite{MILLER20191}, existing methods remain insufficiently human-centred in healthcare settings, where explanation recipients vary in uncertainty comprehension and face significant emotional, financial, and decision-making consequences \cite{ehsan2021expanding} \cite{wang2019designing}.
Healthcare outcome prediction models and risk communication to lay audiences have been studied for decades. However, prediction models increasingly integrate multiple data sources and use complex black-box architectures while simultaneously becoming more lay user (patient) facing. Examples include online prediction tools such as Qrisk tools in the UK and health chatbots\cite{miner2020chatbots}, often used without expert oversight. Expanding the XAI research considerations to address these user needs requires quantifying understandability issues and designing the interfaces for utility and trust. This user study advances this broader research agenda.
The Outcome Prediction in Subfertility Tool (OPIS) \footnote{\url{https://w3.abdn.ac.uk/clsm/opis/}} is a publicly available tool for estimating the probability of live birth following in vitro fertilisation (IVF) at three different stages of treatment. Our study focuses on the OPIS pre-IVF tool, which allows users to input patient characteristics known prior to treatment. It aids risk assessment and expectation management during infertility treatment. OPIS is based on the McLernon model \cite{McLernon2016-tl}, trained on the Human Fertilisation and Embryology Authority (HFEA) data from 113,873 women who underwent complete IVF cycles in the UK between 1999 and 2008 (C index of 0.73).  The tool predicts the cumulative probability of live birth over six IVF cycles, displaying results as a probability graph along with a textual explanation. Figure \ref{fig:opisUI}  shows a screenshot of the tool interface. A higher resolution image is available in the GitHub repo associated with the paper\footnote{\url{https://github.com/adarsa/OPIS-evaluation-phase1}}. 

We evaluate user feedback collected from the tool. The quantitative rating shows a difference in understandability of results between different types of users. Thematic analysis reveals recurring patient queries about \emph{features} that they consider important but not considered for the outcome prediction. XAI research typically explains model reasoning by focusing on feature space and model parameters. We explore this need for explanation further through a survey of current/past IVF patients followed by a semi-structured interview.  We highlight the insufficiency of current XAI considerations to address the misalignment between the model's feature space, derived from years of population-level studies, and the users' desire for personalised predictions that align with their perceived individual circumstances and needs. In conclusion, we discuss the possibility of a conversational user interface(UI) for personalised explanations.

\section{Understandability of OPIS UI - User Feedback Analysis}
The pre-IVF OPIS makes predictions based on a discrete-time logistic regression model that considers factors: the woman's age, infertility duration, treatment type (IVF, ICSI), year, tubal and male factor infertility, unexplained infertility, anovulation, and previous pregnancy \cite{McLernon2016-tl}. While the model has been externally validated with more recent data \cite{10.1093/humrep/dead165}, the interface has not been evaluated or revised. 

The UI and prediction include a feedback survey for qualitative and quantitative user input. Survey versions vary by user role, with patients assessing: \emph{Understandability of results}, \emph{User-friendliness of}:\emph{ Tool description and usage}, \emph{ Entering the data}, \emph{Presentation of results}, while healthcare professionals also evaluate its usefulness in consultations. Responses are anonymous. We analysed 62 feedback responses from OPIS tool users between May 2020 and October 2024 - 48 patients and 14 healthcare professionals. User ratings revealed distinct differences in perceived user-friendliness. While all healthcare professionals found the tool's result presentation clear, 15\% of patients rated it as not user-friendly, and 22\% reported difficulty understanding the results (Table \ref{tab:tool-survey-percentage}). Alongside the ratings, users also provided textual feedback.  Responses containing single words, gibberish, or irrelevant comments were excluded, along with seven responses linked to different versions of the tool. The final analysis includes 31 responses from 22 unique patients.

\begin{table}[!ht]
    \centering
    \caption{Percentage of users who rated the tool as user-friendly (UF) across different criteria, and the predicted results as understandable and interpretable.}
    \label{tab:tool-survey-percentage}
    \renewcommand{\arraystretch}{1} 
    \begin{tabular}{p{0.18\textwidth}p{0.22\textwidth}p{0.15\textwidth}p{0.15\textwidth}p{0.3   \textwidth}}
        \toprule
        \textbf{User Type} & \textbf{UF-Description} & \textbf{UF-Input} & \textbf{UF-Result} & \textbf{Understandability} \\
        \midrule
        Healthcare Professional & 92\% & 77\% & 100\% & 100\% \\
        Patient & 91\% & 89\% & 85\% & 78\% \\
        \bottomrule
    \end{tabular}
\end{table}

\subsection{User Questions as Need for Explanation}

Liao et al.\cite{liao2020questioning} categorised user's need for explanation or difficulty in understanding model outputs by framing explanations as answers to user queries. Based on interviews with 20 user experience designers, they developed the XAI question bank, which has been applied in human-centred design and conversational interfaces for lay users \cite{Slack2023}. We use the XAI question bank as a reference to analyse survey feedback, applying deductive and inductive coding to identify key themes. 
While the XAI question bank focuses on model reasoning and underlying data, OPIS users primarily question tool design and features. For example:  \emph{I have severe endometriosis, the highest stage with cysts. It doesn't ask any questions regarding this as an underlying condition. I assume I shouldn't be relying on the calculator as my percentage would be lower?}. Although the HFEA dataset used for modelling includes \emph{endometriosis}, it was not considered in the model because it was not a significant predictor of live birth. This information is not communicated to users.
Further, the OPIS tool relies on data collected until 2009 (and validated on data until 2016). Advances in treatment and new evidence about outcome predictors have since emerged, which are not captured in data from earlier years. This is particularly relevant for underrepresented patient subgroups: \\ "\emph{It doesn't take into consideration of people with PCOS who rent}[tend] \emph{to get pregnant naturally later in life.}"\\
"\emph{This tool did not seem to provide for someone using IVF due to being in a lesbian relationship - the amount of time trying to conceive is not relevant as we have gone straight to IVF for our first time of trying.}"\\
We codify this recurring theme across the responses as \textbf{\emph{Feature why not considered}}. We observe questions regarding both features considered in modelling but not implied in the UI and features entirely unknown for the model. Addressing this mismatch in the user's mental model and the data model is requisite for understandability and trust.
The XAI question bank category \emph{XAI data} captures questions on the patient data used to develop the model. OPIS tool users often have additional data knowledge based on prior tests or treatment experience, leading to queries beyond the tool's intended scope. (ex:  \emph{Was embryo quality captured in data?}, \emph{It should allow me to choose if I want to transfer one embryo or two.}) We distinguish these and code them as \textbf{\emph{Modeling assumption}}—capturing epistemic choices in tool design. Additionally, we introduce \textbf{\emph{UI issue}} for reported display problems and \textbf{\emph{Other}} for general feedback. Table \ref{tab:tab:OPISfeedbackcodes_} lists all categories and their definitions. 

\begin{longtable}[c]{@{}p{0.15\columnwidth}p{0.25\columnwidth}p{0.6\columnwidth}@{}}
    \caption{Codes and corresponding definitions used for analysing the User feedback. Codes with a preceding tag XAI are the ones derived from the question bank developed by Liao et al.} \label{tab:tab:OPISfeedbackcodes_} \\
    \toprule
    \textbf{ID} &\textbf{Code} & \textbf{Representative explanation question} \\
    \midrule
    \endfirsthead

    \multicolumn{3}{c}%
    {{\bfseries Continued from previous page}} \\
    \toprule
    \textbf{ID} &\textbf{Code} & \textbf{XAI Question} \\
    \midrule
    \endhead

    \midrule
    \multicolumn{3}{r}{{\bfseries Continued on next page}} \\
    \endfoot
    \bottomrule
    \endlastfoot

        XP& XAI performance &How accurate is the IVF outcome prediction model and what patient factors does it consider?\\
    \midrule
    XD& XAI data & What types of patient data were used to create the IVF outcome prediction model?\\
    \midrule
    XO&XAI output &How should I interpret the predicted percentages?\\
    \midrule
    XPH&XAI performance-how & How does the prediction model work?\\
    \midrule
    XW&XAI why & Why did the model give this specific prediction for a patient? \\
    \midrule
    XTIME&XAI other-time & How could the IVF prediction model improve over time?\\
    \midrule
    XTERM&XAI other-term & What does specific technical term shown in the tool mean in simple terms?\\
    \midrule
    XS&XAI other-social& How have other users got on with the model?\\   
    \midrule
    FEATWN&Feature why not considered& I have specifically factor X which is not considered by the model. Can I trust this model prediction?\\   
    \midrule
    MA&Modeling assumption & Expectations from the tool which are not strictly pre-IVF factors.\\
    \midrule
    UI&UI issue& Issue with viewing the prediction tool on the Users particular device.\\  
    \midrule
    OTHER& Other& General sentiment about tool or concerns about the tool, emotional aspects of the treatment.\\   

\end{longtable} 

\subsection{Discussion}
User-friendliness issues primarily stem from UI rendering problems, particularly with the interactive graph on some mobile devices—a major concern as many users rely on mobile access. Users also found selected inputs unclear and suggested making them more explicit. Another key issue is terminology comprehension (\emph{XAI other term}). The definition of \emph{cycle}, crucial for interpreting the graph, is provided in the UI but appears unnoticed or unclear to many users.

Regarding the understandability of results, the user questions aligned with the XAI question bank regarding prediction confidence. Users raise questions about the underlying data (\emph{XAI data}), the model's performance (\emph{XAI performance}), and the output interpretation(\emph{XAI output}). As seen in the quantitative analysis, interpreting the cumulative probability distribution is challenging for users with varied exposure to data and graphs.  Previous studies have also noted patient difficulty in comprehending probabilities as frequency, percentages, or risk difference\cite{zipkin2014evidence}. In addition to numeric representation, graphical, textual and tabular risk communication methods are also well explored in literature\cite{Spiegelhalter_2017}, however, with no conclusive evidence of an appropriate method.

Apart from the  \emph{UI issues}, the most common theme across all criteria is \emph{Feature why not considered}. Users questioned the omission of features they viewed as essential predictors of live birth, affecting their trust in the tool. In the case of factors PCOS and endometriosis, users felt the tool lacked consideration for diverse personal circumstances. Addressing these concerns requires distinguishing between features explainable within the model's scope -those excluded due to developer choices (e.g., endometriosis, male factors) or domain expertise (e.g., prior use of oral contraceptives) - and those beyond the model's knowledge base, such as previous IVF cycles. XAI research has explored the former through techniques like model reasoning, local explanations, and contrastive or counterfactual explanations to answer \emph{what if} questions. However, addressing features outside the model's knowledge base remains largely unexplored.
\\\\
\textbf{The model's explainability is limited because it cannot address queries about features that are not in its feature space.}
\\\\
Miller et al. \cite{MILLER20191} emphasised that explanations should align with the user's mental model. This principle has shaped the XAI question bank and other systems for lay users. However, our study highlights that limiting explainability to the model's feature space fails to address many user concerns, a noted challenge in healthcare AI regulation\cite{REFORM2024}.

In infertility treatment, patients navigate social, financial, and emotional factors, actively gathering information and comparing IVF prediction tools. Their questions often stem from distrust, as certain features they consider crucial are missing. A framework to address these requires viewing user concerns from the model's perspective, recognising that its predictions are shaped by developer-chosen features and dataset distributions. We propose categorising feature questions into four distinct categories detailed in Table \ref{tab:feature_modelling}. To validate these and gather a more structured understanding of user questions, we further explore the \textbf{research question: What are the questions potential users expect to be answered in a possible explanation of the predicted outcome?}

Analysis of the user response reveals that most user concerns align with questioning the AI framework. Hence, a dialogue-based interface and narrative explanations could be promising solutions. However, further investigation is needed to understand user expectations for such textual or conversational interfaces. \textbf{Research question: What are the user's expectations and concerns regarding a dialogue system for risk communication?} We design a user study to address these questions and quantify the baseline understandability of the current OPIS UI for future work.

\begin{table}[htbp]
    \centering
    \caption{Categories of explanation consideration to address the user's mental model of outcome prediction. }
    \label{tab:feature_modelling}
    \renewcommand{\arraystretch}{1.3} 
    \setlength{\tabcolsep}{2pt} 
    \begin{tabular}{@{}p{0.27\columnwidth}p{0.7\columnwidth}@{}}
        \toprule
         & \textbf{Description}  \\
        \midrule
        Feature has no impact on model & Already considered in modelling (Not explicitly stated in the UI) \\
        Feature excluded by domain knowledge & Earlier studies have shown that it is not an important predictor of outcome. \\
        Feature value out-of-distribution & This factor is known to have an influence on the outcome, but the available data in training is limited.\\
        Unknown feature & This feature is not considered in modelling. For example, its impact is not known when the model was built. \\
        \bottomrule
    \end{tabular}
\end{table}

\section{User Study with Potential IVF Patients}
\subsection{Methodology}
The study consists of an online survey and follow-up interviews with selected participants recruited via fertility support groups in the UK. It has three parts:\\
\textbf{UI Evaluation (Survey Part 1)}: Participants assess the current OPIS UI on understandability and trust. Ratings include perceived understandability (yes/no/partially) and a 3-point trust scale \cite{hoffman2021measuring}. Participants also answer comprehension questions based on the graph output to measure understandability. Responses are anonymous, and demographic data, including numeracy levels \cite{weller2013development}, are collected.\\
\textbf{Explainability Needs (Survey Part 2)}: Participants review an adapted XAI question bank for IVF and select questions that best reflect their explanation needs. They also submit additional concerns about model trust and reasoning. Responses are analysed using codes developed using previous user feedback.\\
\textbf{Semi-Structured Interviews (Part 3)}: Four participants from the survey, selected based on prior experience with prediction tools or patient support groups, provide deeper insights into explainability needs. Topics include model accuracy, probability interpretation, and expectations for a dialogue-based interface (RQ4). No medical data is collected, and interview responses are manually analysed for data security. Details of the survey and interview questions are in Appendix.

\subsection{Evaluation of Results}
Thirteen participants took the survey. We excluded two participants who reported to have no experience with IVF. The survey averaged 10 minutes to complete. Lower participation was due to mobile compatibility issues and a potentially unfamiliar format, which limited accessibility. Table \ref{tab:demographics} summarises the demographics of the participants.

\begin{table}[!]
    \centering
    \caption{.}
    \label{tab:demographics}
    \renewcommand{\arraystretch}{1} 
    \begin{tabular}{@{}p{0.3\columnwidth}p{0.65\columnwidth}@{}}
        \toprule
        \textbf{Category} & \textbf{Details (Count)} \\
        \midrule
        Gender & Male (2), Female (9) \\
        Country of residence & UK (9), Spain (1), India (1) \\
        Identifying as LGBTQ+ & No (8), Yes (2), Prefer not to say (1) \\
        Highest Education level & Master or PhD Degree (8), Bachelor Degree (3) \\
        Rasch numeracy level & High(3), Average to above-average(3), Below average(2), Low(3)\\
        \bottomrule
    \end{tabular}
\end{table}

The normalised average user perceived understandability(0.86) is slightly higher and negatively correlated (spearman coefficient:-0.4) to the task-based score of 0.76. User-perceived understandability has a strong positive correlation(0.75) to the trust score. Since the number of participants is low, we do not report a demography specific analysis. 

We evaluate the text feedback from the survey and interview using the same code developed in Table \ref{tab:tab:OPISfeedbackcodes_}. 19 feedbacks were collected from the survey participants across the four criteria: trust, understandability, need for explanation, and overall feedback. For the interview, we consolidated the responses from all questions, and the first author coded the parts relevant to the explanation requirement. We present representative excerpts from all the user responses under each theme in Table \ref{tab:theme_response_example}. 

\begin{longtable}[c]{@{}p{0.15\columnwidth}p{0.85\columnwidth}@{}}
    \caption{Representative excerpts of participant responses.} \label{tab:theme_response_example} \\
    \toprule
    \textbf{Category} & \textbf{Survey and interview excerpts} \\
    \midrule
    \endfirsthead

    \multicolumn{2}{c}%
    {{\bfseries Continued from previous page}} \\
    \toprule
    \textbf{Category} & \textbf{Survey and interview excerpts} \\
    \midrule
    \endhead

    \midrule
    \multicolumn{2}{r}{{\bfseries Continued on next page}} \\
    \endfoot
    \bottomrule
    \endlastfoot
        XP& - What was the model accuracy after the evaluation? \\
        & - ... these tool work well for you know  standard case. For cases with lesser data, it would be complicated.\\
        \midrule
        XD  & - I like to know how you collected particular type of data based on different patient, different age factors with the same patient factors. \\
        & - I suppose perhaps the more complicated you are the harder it must be those tool, perhaps less data in smaller pool. \\
        \midrule
        FEATWN & - We don't know what PGT-A tests will show, etc.  All of these factors are not being taken into account.  \\
        & - Can you include the likelihood when combining with taking certain advised medication before / during? Is past history of pregnancy complications taken into consideration.\\
        & - There is environmental factors a smoking, alcohol... There are male factors ..  quality of sperm..not captured. \\
        & - Endometrosis or PCOS patients how (are these) factored? \\
        & - It should be more personalised...Obesity is quite a big factor in whether IVF is successful or not.\\
        \midrule
        MA & - Does it matter if it is 6 separate egg collections or if all transfers are from embryos resulting from one collection? \\
        & ..or example, it could include the statistics for unassisted cycles.\\
        \midrule
        XO & - Why the percentage chance is higher after each cycle? \\
        &  - It is upsetting for people who believe they are on "the wrong side of the statistics". I have had 7 transfers .. (but I) hear numbers like 83\% success rate. \\
        \midrule
        XW & - Perhaps some sort of visual way of highlighting reasons behind prediction eg factors age / previous pregnancy / tube issues. \\
        & - Different ovulation problems will have a different potential outcome; we don't know what PGT-A tests will show, etc.\\
        \midrule
        XTERM & - It would be good to clarify the definition of cycle - I am interpreting it as one collection and one single embryo transfer?  \\
        & - There's a lot of confusion with the definition of cycle.   \\
        \midrule
        XPH & - So this graph shows all the time it is a positive way...What about the negative factors, anything that are responsible for the IVF.\\
    
\end{longtable}
 \begin{figure}[h]
\centering
\includegraphics[width=\linewidth]{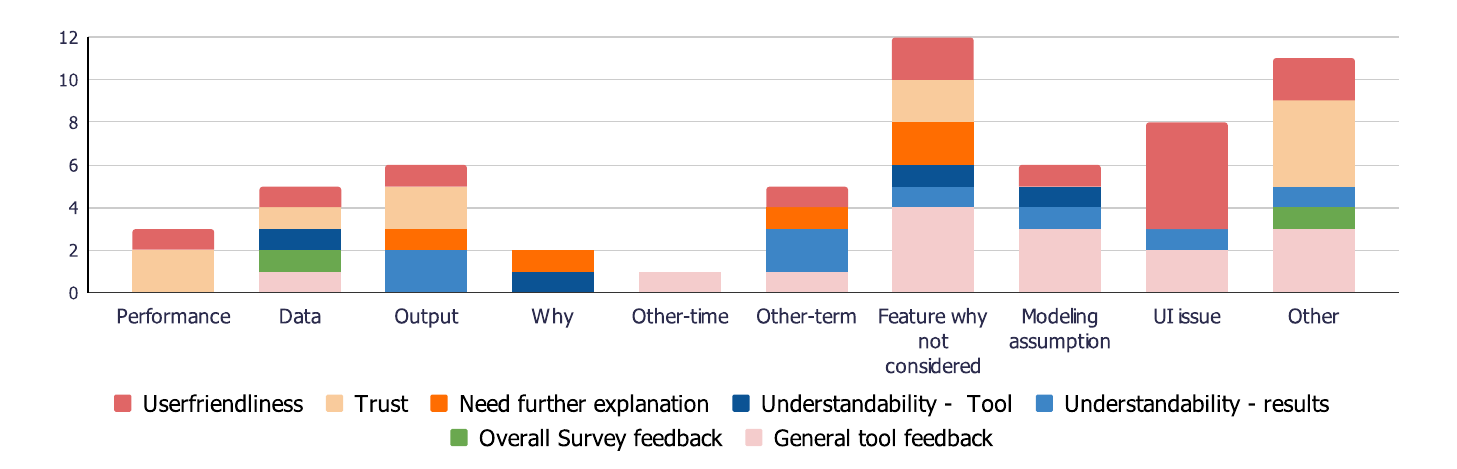}
\caption{A summary of coded user intents from the tool feedback, survey part 1.}
\label{fig:summary}
\end{figure}
Under \emph{FEATWN}, we further see multiple user queries falling under the identified categories. Although many participants expressed concerns about the underlying data and prediction confidence, only one mentioned model accuracy. However, when asked to select an XAI intent from a predefined list, 8 out of 11 chose XAI performance or XAI data. Similarly, no survey participants explicitly questioned model reasoning, yet 6 selected it from a dropdown. This suggests that lay users may struggle to articulate concerns about model validity in technical terms and may raise it as a related question. Additionally, no participants raised concerns about the tool’s applicability to others unless explicitly prompted during interviews, reinforcing that their primary focus is on personal relevance and clarity rather than general validity. Five responses are coded as \emph{other}, highlighting the emotional and financial impact of IVF predictions. Unlike general user feedback, the survey underscores the emotional and financial dimension of the treatment. All the interview participants also mention this aspect.  Figure \ref{fig:summary} summarises the distribution of queries across different categories from the user feedback, parts 1 and 2 of the survey.

\subsection{Dialogue Interface - User Expectations and Future Work}
Integrating human awareness into the tool through an iterative \emph{dialogue} can adapt explanations to user's needs and concerns. In user interviews, all participants were familiar with chatbots. Two responded positively, with one noting its potential usefulness in the UK, where accessing fertility specialists NHS may be lengthy. However, one participant who is experienced with chatbots also mentioned frustration when they fail to provide personalised responses and emphasised the need to understand user acronyms. Other concerns included \emph{privacy, accessibility,} and \emph{language clarity}.
Interactive, narration-based natural language generation has been explored for XAI to personalise explanations and communicate uncertainty. Building on the XAI question bank, user studies have suggested leveraging LLMs to interpret user needs and translate specific information into layperson-friendly language as a promising approach\cite{liao2020questioning,Slack2023}. However, we highlight the need for such systems to detect user intents beyond strict model feature spaces. 

Human-AI interaction frameworks, such as argumentation-based dialogue \cite{sklar2018explanation}, belief-desire-intention models \cite{dennis2022explaining} could enhance human-awareness in health informatics and our findings contribute to achieving this. Interactive, narration-based explanations have been noted to support the personalisation of explanations, particularly for lay users \cite{reiter-2019-natural}. Large language models enable this approach through coherent language generation; however, limitations in reasoning, prompt sensitivity, and hallucinations make general-purpose systems unsuitable. Retrieval Augmented Generation (RAG)\cite{lewis2020retrieval} systems offer a promising solution by grounding LLMs in factual, curated data. Recent research in dialogue system evaluation frameworks focusing on clinically meaningful axes is a promising evaluation approach \cite {tu2025towards}. Formalising the user needs in this study will support exploring dialogue-based explanations. Work on a dialogue-based explainer is currently underway.

\section{Related Work}

For a patient or general public, the impact of the method and modality of communicating statistics and their dependence on audience demography has been long studied in risk communication \cite{fischhoff1993risk,Peters2008-dp,Spiegelhalter_2017}. With computer-aided patient communication,  the earliest studies highlight the importance of personalization\cite{jones1999randomised} and the positive effect of underlying model explanations on trust. As models become more complex and opaque, the need to iterate the paradigms for interpretability in decision support systems is seen across different applications\cite{Xu2023-hz}. Xu et al. highlight the lack of consensus in evaluation systems for interpretability and the need to focus on user concerns in clinical model explainability.  In IVF outcome prediction, text and visual interfaces have been shown to impact a better risk understanding \cite{AUstraliaIVF}.  However, conveying the model reasoning and the causal relationships between input and output fall within the scope of explainability research.

We have seen how this is a problem beyond the typical assumptions of closed-world models. The need for models to account for knowledge limitations is highlighted by Phillips et al. \cite{phillips2021four}. Inadequacy of standard performance metrics in capturing model shift of clinical risk prediction has been shown in Cabanillas et al.\cite{cabanillas2024longitudinal}. Recent research has focused on providing pragmatic frameworks for discussing such explanation needs\cite{nyrup2022explanatory}, bridging the gap in model development and adoption \cite{markowetz2024all} and enforcing regulatory requirements\cite{REFORM2024,stoger2021medical}. This study is a step in that direction. 

\section{Conclusion}
In this study, we evaluated an online prediction tool actively used by people planning their IVF. We have identified the unique explanation design considerations for lay users by involving people with lived experience of infertility treatment.
A key takeaway from our analysis is that the current approach to model explainability in  AI research cannot be applied to a patient-facing setting. 
In particular, we have highlighted two user needs. Firstly, we note that addressing a lay user's need for prediction model explainability requires methods of acknowledging out-of-distribution questions on model reliability beyond its feature space. 
Secondly, in the IVF outcome prediction scenario, user questions of model validity are specific to their medical conditions. This need for personalisation has been noted in other domains, and we propose interactive explanations to address them. We have gathered preliminary user feedback for a dialogue-based UI and aim to explore concrete paradigms in future work.

\begin{credits}
\subsubsection{\ackname} We thank the user study participants for their contributions. We also thank the anonymous reviewers for their feedback, which has helped improve this work. A. Sivaprasad's PhD is part of the NL4XAI project, which has received funding from the European Union’s Horizon 2020 research and innovation programme under the Marie Skłodowska-Curie Grant Agreement No. 860621.

\subsubsection{\discintname}
This study protocol was reviewed and approved by The Physical Sciences and Engineering Ethics Board, University of Aberdeen.  The data collected as feedback and surveys were anonymised at the source, and personal information in the interview transcripts was manually removed and annotated by the lead author.

\end{credits}

\bibliographystyle{splncs04}
\bibliography{mainaiih}
%
%
%
%
\appendix

\section{Appendix}
\begin{table*}[]
\renewcommand{\arraystretch}{1.2}
  \caption{Question from the user feedback form in the OPIS tool.}
  \label{tab:OPISquestions}
  \renewcommand{\arraystretch}{1} 
  \begin{tabular}{p{0.03\textwidth}p{0.6\textwidth}p{0.37\textwidth}}
    \toprule
    \textbf{ID} & \textbf{Question Text} & \textbf{Question Text Code} \\
    \midrule
    Q1 & Are you a healthcare professional or a patient? & OPIS-USER-TYPE \\
    Q2 & Have you used OPIS in a consultation with couples who were about to begin IVF? & OPIS-HP-USEWITHPATIENT \\
    Q3a & Did you think OPIS was user friendly with regards to: The wording used to describe the tool and how to use it? & OPIS-HP-USERFRIENDLY-DESCRIPTION \\
    Q3b & Did you think OPIS was user friendly with regards to: How you entered patient data? & OPIS-HP-USERFRIENDLY-INPUT \\
    Q3c & Did you think OPIS was user friendly with regards to: The presentation of results i.e. the graph with supporting text? & OPIS-HP-USERFRIENDLY-RESULT \\
    Q3d & Did you think OPIS was user friendly with regards to: If you wish to explain your answers, please provide feedback with any suggestions for improvement. & OPIS-HP-USERFRIENDLY-FEEDBACK \\
    Q4a & Were you able to understand and interpret the results? & OPIS-HP-UNDERSTANDABILITY \\
    Q4b & Were you able to understand and interpret the results? If you wish to explain your answer, please provide feedback with any suggestions for improvement. & OPIS-HP-UNDERSTANDABILITY-FEEDBACK \\
    Q5a & Did OPIS help support your consultation with your patient(s)? & OPIS-HP-USEFULL-CONSULTATION \\
    Q5b & Did OPIS help support your consultation with your patient(s)? : Please explain further. & OPIS-HP-USEFULL-CONSULTATION-FEEDBACK \\
    Q6a & Would you use OPIS again? & OPIS-HP-USEFULL-REUSE \\
    Q7 & If you have any additional comments, please include below: & OPIS-HP-FEEDBACK \\
    Q8a & Did you think OPIS was user friendly with regards to: The wording used to describe the tool and how to use it? & OPIS-PATIENT-USERFRIENDLY-DESCRIPTION \\
    Q8b & Did you think OPIS was user friendly with regards to: How you entered your data? & OPIS-PATIENT-USERFRIENDLY-INPUT \\
    Q8c & Did you think OPIS was user friendly with regards to: The presentation of results i.e. the graph with supporting text? & OPIS-PATIENT-USERFRIENDLY-RESULT \\
    Q8d & Did you think OPIS was user friendly with regards to: If you wish to explain your answers, please provide feedback with any suggestions for improvement. & OPIS-PATIENT-USERFRIENDLY-FEEDBACK \\
    Q9a & Did you understand what your results meant? & OPIS-PATIENT-UNDERSTANDABILITY \\
    Q9b & Did you understand what your results meant?: If you wish to explain your answer, please provide feedback with any suggestions for improvement. & OPIS-PATIENT-UNDERSTANDABILITY-FEEDBACK \\
    Q10 & If you have any additional comments, please include below: & OPIS-PATIENT-FEEDBACK \\
    \bottomrule
  \end{tabular}
\end{table*}

\begin{table}[htbp]
    \centering
    \caption{Online User Survey Questions}
    \label{tab:survey_questions}
    \renewcommand{\arraystretch}{1.5} 
    \begin{tabular}{p{0.2\textwidth}p{0.65\textwidth}}
        \toprule
        \textbf{Question Name} & \textbf{Question} \\
        \midrule
        understand\_yes\_no & For Anna, the AI model predicts the chance of live birth as shown in the Graph. Do you understand what these results meant? \\
        understand\_text & Which parts do you not understand? \\
        understand\_expect & What additional explanation of the prediction would make it more understandable? \\
        understand\_eval\_1 & Based on the prediction tool, what is Anna's expected chance of success after the first cycle of IVF? \\
        understand\_eval\_2 & Based on the graph, what is the highest chance of live birth after 3 cycles? \\
        understand\_eval\_3 & Based on the graph, what is Anna's percentage chance of success after 6 complete cycles of IVF? \\
        trust\_1 & To what extent do you agree with the following statements? I have a realistic view of what would be the outcome for this patient. \\
        trust\_2 & To what extent do you agree with the following statements? I approach this tool with caution. \\
        trust\_3 & To what extent do you agree with the following statements? How likely are you to recommend this tool to a friend? \\
        trust\_text & Do you wish to explain any of the answers above? \\
        XAI\_intent & Let us re-visit patient Anna and the OPIS prediction. You may have questions about how this tool works or the trustworthiness of its predictions. What are your concerns about this tool? To help you, below are questions that other users have asked. Please choose the questions you think are important from this list. You may select multiple options. \\
        XAI\_text & Do you have any questions/queries other than the ones above? \\
        feedback & Please note any additional remarks. \\
        \bottomrule
    \end{tabular}
\end{table}

\begin{table*}[]
  \caption{Questions for the semi-structured interview.}
  \label{tab:InterviewQuestions}
  \renewcommand{\arraystretch}{1.5} 
  \begin{tabular}{p{0.24\textwidth}p{0.4\textwidth}p{0.35\textwidth}}
    \toprule
    \textbf{Category} & \textbf{Question } & \textbf{Probing question} \\
    \midrule
    \textbf{BACKGROUND INFORMATION} & Could You briefly describe your familiarity with IVF outcome prediction tools? &  \\
     & How would you describe your experience in interpreting data and probability? &  \\
     & What aspects of the OPIS tool stood out to you? &  \\
     \textbf{CONCERNS ABOUT MODEL ACCURACY AND VALIDITY} & Do you have concerns about the accuracy of the prediction model? & Could you describe any specific aspects of the model’s predictions that you find unreliable? \\
     & &Do you have questions about the validity of the model across different subgroups of patients, such as age or health conditions?\\
     \textbf{UNDERSTANDING AND INTERPRETING PREDICTED PROBABILITIES}&What questions do you have about interpreting the predicted percentages the model presents?&What will make these predictions more understandable?\\
     &Based on your experience, what challenges do people commonly face when interpreting probabilities or percentages?&How do you think this applies to interpreting the IVF model’s predictions?\\
     \textbf{PERCEPTION OF HOW THE MODEL MAKES PREDICTIONS}&How do you think the AI model makes the predictions?&\\
     &What patient features or factors do you think are the most important predictors of IVF success in this model?&In your opinion, how do specific factors related to each patient affect the model’s predictions?\\
     &Are there additional factors you think should be considered?&Why do you think these factors are important?\\
     & &In your opinion, how could including these additional affect the validity of the prediction?\\
     \textbf{DIALOGUE}&If we are to design a dialogue interface for OPIS, what are your expectations?&Do you have any concerns about such an interface?\\
     \textbf{CLOSING REMARKS}&What are your suggestions for improving the tool?&\\
     &What is your feedback on this interview and the survey?&\\
    \bottomrule
  \end{tabular}
\end{table*}

\end{document}